# Superdirectional Beam of Surface Spin Wave

A. Yu. Annenkov, S. V. Gerus and E. H. Lock*

*Kotel'nikov Institute of Radio Engineering and Electronics of Russian Academy of Sciences, Fryazino branch*

The visualized diffraction patterns of surface spin wave excited by arbitrarily oriented linear transducer are investigated experimentally in the plane of tangentially magnetized ferrite film for the case where the transducer length $D$ is much larger than the wavelength $\lambda_0$. It is shown experimentally and theoretically that the angular width of diffracted surface spin wave beam in anisotropic ferrite film can take values greater or less than $\lambda_0/D$ and can also be zero. For the last case superdirectional (nonexpanding) beam of the surface spin wave is observed experimentally: the smearing of the beam energy along the film plane is absent and the length of the beam trajectory is maximal (~50 mm). It is found, that well known Rayleigh criterion used in isotropic media can't be used to estimate the angular width of spin wave beams. The experimental results are in good agreement with theoretical investigations, predictions, calculations and formulas obtained recently.

Recently it was investigated theoretically a two-dimensional diffraction patterns arising in the far-field region of a ferrite film surface for the common geometry, when a plane surface spin wave with noncollinear group and phase velocities is incident on the wide slit in opaque screen with arbitrary orientation [1, 2]. In contrast to the similar problem for isotropic media it was found in magnetostatic approximation [3], that the angular width $\Delta\psi$ of the diffracted spin wave beam is defined not only by the ratio $\lambda/D$ (here $\lambda$ is the incident wavelength, and $D$ is the slit length) but also by the mathematical properties of isofrequency dependence (ID) for the diffracted wave (in other works ID can be termed as "section of wavevector surface", "section of the isoenergy surface", "equifrequency line", ets. [2]). The general formula for the angular width $\Delta\psi$ of a diffracted beam was derived as a result of this investigation. It was shown that this formula is valid not only for various types of spin waves, but also for other waves, propagating in various anisotropic media and structures (including metamaterials, which are characterized by ID too). As follows from the obtained formula, in contrast with Rayleigh criterion the angular width of the beam in anisotropic media can not only take values greater or less than $\lambda/D$, but also can be equal to zero under certain conditions [2]. Analysing the mathematical properties of the wave ID in various anisotropic media and structures, one can find out specific geometries giving possibility to excite experimentally these superdirectional beams (i.e. beams with zero angular width). Using the same analysis it is found in the works [4, 5] that superdirectional beams of the backward spin waves can be observed in the free ferrite slab too. Mention must be made, that the similar results for the backward spin wave beams was predicted also recently in the paper [6], where for superdirectional propagation of the beam it was used the terms "focusing" and "caustic propagation".

In this Paper, we demonstrate the validity and availability of the general formula described above and discuss below experimental diffraction patterns of the surface spin wave beams propagated in a tangentially magnetized ferrite film for certain geometries, including the case where the angular width of the beam can be equal to zero.

The general scheme of experimental setup is shown in Fig. 1. Propagation of spin wave beams is studied in yttrium iron garnet (YIG) film grown on a 0.5-mm thick substrate of gadolinium gallium garnet (GGG). YIG film (*5* in Fig. 1) having diameter 76 mm, thickness $s = 14.7$ μm and magnetization $4\pi M_0 = 1855.8$ Gs is magnetized to saturation by the tangential uniform magnetic field $\mathbf{H}_0$ with a value $H_0 = 471.5$ Oe. Microwave signal, propagating as a spin wave in the YIG film from input transducer *6* to output transducer *4*, is generated and received by device *1*, that can measure also a complex coefficient of transmission signal. Analog signal received by device *1* is transformed then by analog-to-digital converter and is analysed by computer *2*. Input transducer *6*, exciting spin wave beam, and output receiving transducer *4* are made of gold-plated tungsten wire of 12 μm thick. Exciting transducer *6* has the length $D = 5$ mm and receiving transducer *4* has the aperture $L \sim 0.5$ mm. Two identical systems provide in-plane displacement of the receiving transducer in two orthogonally directions of YIG film plane (along *y* and *z* axes). Thus, receiving transducer equipped also by the special position sensors *3* is turned into movable scanning probe, directed by computer *2*.

The distribution of spin wave beam in the plane of the yttrium iron garnet film is studied experimentally by means of the probe scanning method, that was used earlier in [7, 8]. Recently this method was improved cardinally [9] and now it is possible to visualise distribution of amplitude and phase of spin wave along the film surface by means of computer signal processing (for more detail see [9]). Note that dipole spin wave distribution may occupy a large area compared with YIG film surface of 76 mm in diameter. Therefore, the described measurement method is appear much more effective for dipole spin waves than the similar well known method based on the Brillouin scattering of light by the spin wave [10] (it would take a lot of time to probe the whole YIG film surface by the Brillouin method).

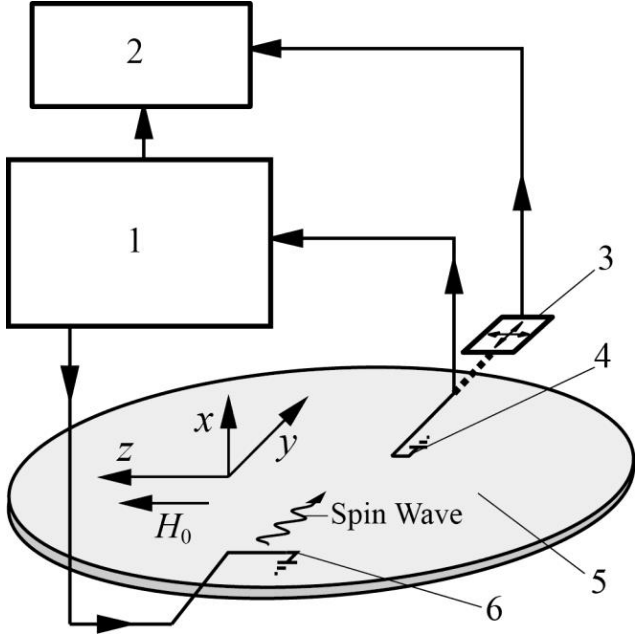

Fig. 1. Schematic view of the experimental setup: *1* – device measuring complex coefficient of transmission signal, *2* – computer, *3* – sensors of the transducer position, *4* – receiving transducer (probe), *5* – YIG film, *6* – linear transducer exciting spin wave.

Mention must be made that theory [2] is also valid for the case, where spin wave beam is excited by arbitrary oriented linear transducer. In this case, one can assume that the spin wave vector $\mathbf{k_0}$ is approximately perpendicular to the linear exciting transducer (this assumption is justified in section 9 of [2]) and, therefore, general formula (30) in [2], describing angular beam width $\Delta\psi$, is simplified to the next form

$$\Delta\psi = \frac{\lambda_0}{D}\left|\frac{d\psi}{d\varphi}(\varphi_0)\right|. \qquad (1)$$

Here angle $\varphi_0$ is orientation of the wave vector $\mathbf{k_0}$ respect to the *y* axis (or angle $\varphi_0$ may be considered approximately as orientation of transducer line respect to the vector $\mathbf{H_0}$); angles $\varphi$ and $\psi$ are orientations for an arbitrary wave vector $\mathbf{k}$ and a corresponding group velocity vector $\mathbf{V}$ respectively; $d\psi/d\varphi(\varphi_0)$ – the derivative value, corresponding to the point of ID in which the wave vector $\mathbf{k_0}$ is ended (for more detail see section 8 in [2]).

Characterizing spin wave beam parameters it is convenient to use both *absolute* angular beam width $\Delta\psi$ and *relative* angular beam width $\sigma$ which is equal to the ratio between the value $\Delta\psi$ (in radians) and the value $\Delta\psi_{isotr} = \lambda_0/D$ representing the angular width of a similar diffracted beam in isotropic medium (in radians too):

$$\sigma = \frac{\Delta\psi}{\Delta\psi_{isotr}} = \frac{\Delta\psi}{\lambda_0/D} = \left|\frac{d\psi}{d\varphi}(\varphi_0)\right|. \qquad (2)$$

From the physical standpoint, relative angular width $\sigma$ shows how much value $\Delta\psi$ is smaller (or greater) than the absolute angular width $\Delta\psi_{isotr}$ for the similar beam (i.e., for the beam with the same ratio $\lambda_0/D$) propagating in isotropic medium.

Formula (1) shows, that the angular beam width $\Delta\psi$ is determined not only by the ratio $\lambda_0/D$, but also by the derivative value $d\psi/d\varphi(\varphi_0)$ characterizing the curvature of ID. So if one can find the structure (medium) and geometry (excitation topology), at which $d\psi/d\varphi(\varphi_0) = 0$, then one can observe superdirectional (nonexpanding) beam with angular width $\Delta\psi = 0$, i.e. the beam retaining constant absolute width during its propagation!

As it is found in [2], the value $d\psi/d\varphi(\varphi_0)$ can be equal to zero for surface spin wave with frequency lying not far from the beginning frequency of the spectrum (for more detail see Fig. 7b, curve 2 in [2])

$$f_{beg} = \gamma\sqrt{H_0(H_0 + 4\pi M_0)}/2\pi. \qquad (3)$$

So to observe superdirectional (nonexpanding) beam corresponding to the case $d\psi/d\varphi(\varphi_0) = 0$ we have calculated ID for the surface spin waves with various frequencies (Fig. 2) and corresponding dependences $\psi(\varphi)$ (Fig. 3a), $d\psi/d\varphi(\varphi)$ and $\sigma(\varphi)$ (Fig. 3b) for the parameters of YIG film mentioned above. Since it is shown in [2] that numerical calculations of value $\sigma$ and calculations based on formula (2) coincide with good accuracy (see Fig. 8 – 11 in [2]), so only values $\sigma$ calculated by the formula (2) are given in Fig. 3b.

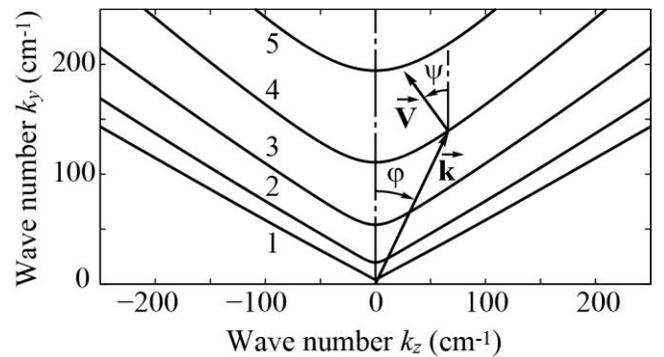

Fig. 2. Isofrequency dependences of the surface spin wave for the frequencies 2950, 2999, 3100, 3242 and 3400 MHz (curves 1 – 5 respectively). An arbitrary vectors $\mathbf{k}$ and $\mathbf{V}$ with their orientation angles $\varphi$ and $\psi$ are also shown.

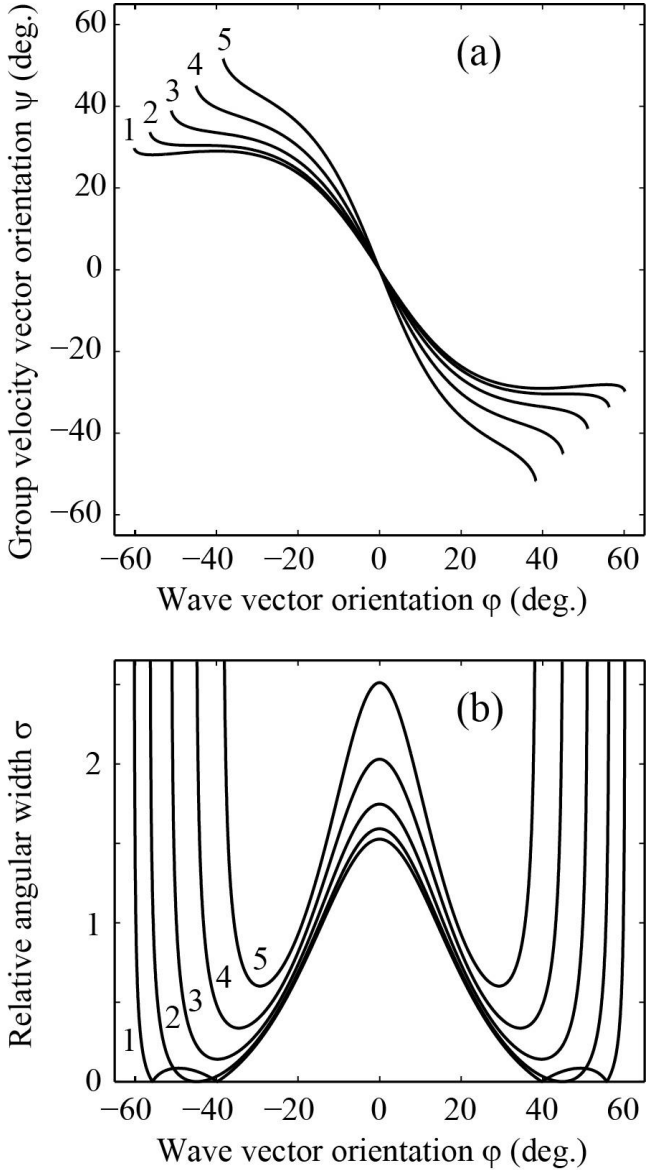

Fig. 3. Orientation angle $\psi$ of the group velocity vector **V** (a) and relative angular width $\sigma$ (b) as a function of the orientation angle $\varphi$ between the wavevector **k** of the surface spin wave and the *y*-axis for the frequencies 2950, 2999, 3100, 3242 and 3400 MHz (curves 1 – 5 respectively).

Then to observe experimentally superdirectional beam propagation, we fixed position of exciting transducer at an orientation $\varphi_0 = -45°$ for which the relative angular width $\sigma$ can be zero for frequency about 3 GHz (see Fig. 3b, curve 2). Since calculations based on the theory [2] do not take into account a number of parameters (such as anisotropy, applied field and magnetization inhomogeneity, etc.) influencing on the spin waves characteristics, visualized patterns in the 120 MHz frequency band with a 1 MHz step are obtained for spin wave beams propagation at this orientation $\varphi_0$. Below in Fig. 4 we present the measured visualized pattern having the minimal angular beam width for the spin wave frequency $f_0 = 2999$ MHz. For the selected values $f_0$ and $\varphi_0$ the spin wave vector $\mathbf{k_0}$ had magnitude $k_0 = 56.5$ cm$^{-1}$ (wave length $\lambda_0 = 1110$ μm) and the ratio $\lambda_0/D$ was 0.222.

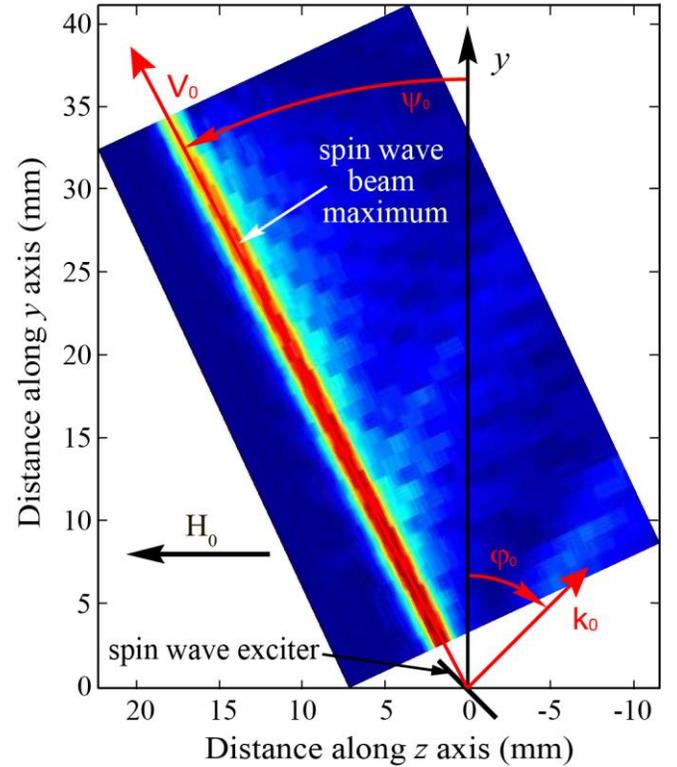

Fig. 4. Amplitude distribution of superdirectional spin wave beam in the plane of ferrite film for the next beam parameters: $f_0 = 2999$ MHz, $k_0 = 56.5$ cm$^{-1}$, $\lambda_0 = 1110$ μm, $\lambda_0/D = 0.222$, $\varphi_0 = -45°$, $\Delta\psi_{exp} = 0.4°$, $\sigma_{exp} = 0.03$. Colour or greyscale change corresponds to the 3 dB change of spin wave amplitude relative to the maximal amplitude.

The pattern in Fig. 4 describes the distribution of spin wave amplitude in the YIG film plane. As it is seen from Fig. 4 surface spin wave has noncollinear orientation of the wave vector $\mathbf{k_0}$ and group velocity vector $\mathbf{V_0}$. It should be noted, that in isotropic media such beam (having ratio $\lambda_0/D = 0.222$) would have angular width about 12° whereas in Fig. 4 spin wave beam has absolute angular width $\Delta\psi_{exp} = 0.4°$ and relative angular width $\sigma_{exp} = 0.03$! These experimentally measured values $\Delta\psi_{exp}$ and $\sigma_{exp}$ are slightly differ from calculated zero value. This fact is explained as follows: since the aperture of the receiving probe is 0.5 mm and the maximal length of the spin wave trajectory is ~ 50 mm (because of the presence of dissipation in real YIG films), the minimal angle that can be measured is ~ 0.5/50/2 = 0.005 rad $\approx$ 0.3°. Unfortunately, we have to state that since spin waves, in contrast to light waves, propagate only at a distance ~ 50 mm, so it is possible to measure only the angular beam width exceeding 0.3° by means of a probe with aperture 0.5 mm. Mention must be made that the use of the probe with a smaller aperture does not increase the accuracy of measurement at the large distances because the probe sensitivity is decreased proportionally to the aperture size. We also note that, since the superdirectional beam does not expand practically, the orientation $\psi_0$ of the group velocity vector for this beam can be measured with the highest accuracy.

In order to demonstrate how strongly the relative angular beam width $\sigma$ can change in anisotropic media, we study also visualized picture of the surface spin wave beam propagation for the geometry, where relative angular beam width $\sigma$ has high value ~ 2. As it may seem from Fig. 3b, geometries characterized by the value $\sigma \sim 2$ can be realized at angles $\varphi_0$ lying near angles of cutoff $\varphi_{cut}$, but these geometries can't be realized because it is impossible to excite spin waves with very large value of the wave number $k_0$. However, value $\sigma \sim 2$ appears also in the geometries, where surface spin wave is excited by linear transducer oriented closely to the vector $\mathbf{H_0}$, i.e. at $\varphi_0 \sim 0$ (Fig. 3b). Below we describe the beam propagation for the most widely used geometry where transducer is orientated along vector $\mathbf{H_0}$. Superposition of both amplitude and phase distributions for this geometry is shown in Fig. 5 for spin wave frequency $f_0 = 3242$ MHz, $k_0 = 110.8$ cm$^{-1}$ ($\lambda_0 = 567$ μm) and $\lambda_0/D = 0.113$.

As one can see from Fig. 5, maximal part of the spin wave energy is transferred along the group velocity vector $\mathbf{V_0}$ in the form of the gradual widening main beam, having angular width $\Delta\psi_{exp} = 12.9°$ (in the Fig. 5 brightest area located along $y$ axis corresponds to the main beam). The experimental value $\Delta\psi_{exp}$ is measured for the part of the beam lying at interval $y < 22$ mm. At $y > 22$ mm amplitude of the beam is small and it is hardly to analyze amplitude distribution in the plane of ferrite film with good accuracy. Note, that little part of the spin wave energy spreads on the whole rest of area located between two straight lines corresponding to the cut-off angles $\psi_{1cut}$ and $\psi_{2cut}$ of the group velocity (Fig. 5). Large deep areas located below these straight lines correspond to the ferrite film regions in which spin wave can't exist. Most dark narrow angle sectors in the Fig. 5 located along the angles, at which magnetic potential of the spin wave beam is equal to zero (amplitude distribution for magnetic potential of studied spin wave beam in the far-field region is similar to the distribution shown in the Fig. 5 (curve 1) in [2]). On the contrary, brightest area near the $y$ axis in the Fig. 5 corresponds to the greatest maximum of the magnetic potential distribution.

It is usually assumed that if the exciting linear transducer is oriented parallel to the vector $\mathbf{H_0}$, a surface spin wave with collinear orientation of the vectors $\mathbf{k}$ and $\mathbf{V}$ is excited in a ferrite film and both $\mathbf{k}$ and $\mathbf{V}$ vectors are directed exactly along the optical axis $y$ (named also axis of collinear propagation). However, as it is seen from Fig. 5, the spin wave fronts are slightly inclined respect to the transducer line and the beam path (location of the greatest beam maximum) is only approximately directed along $y$ axis – in fact inclination of the beam path to the $y$ axis is about 5°. This inclination arises because of at ~3 GHz (wavelength ~10 cm) there is a small phase difference between the initial and final points of exciting linear transducer (distance between these points is 5 mm) and this phase difference leads to the certain inclinations for both wave vector $\mathbf{k}$ and group velocity vector $\mathbf{V}$ from the normal to the exciting transducer line. Thus, exactly speaking, linear transducer is not cophased exciter at microwaves.

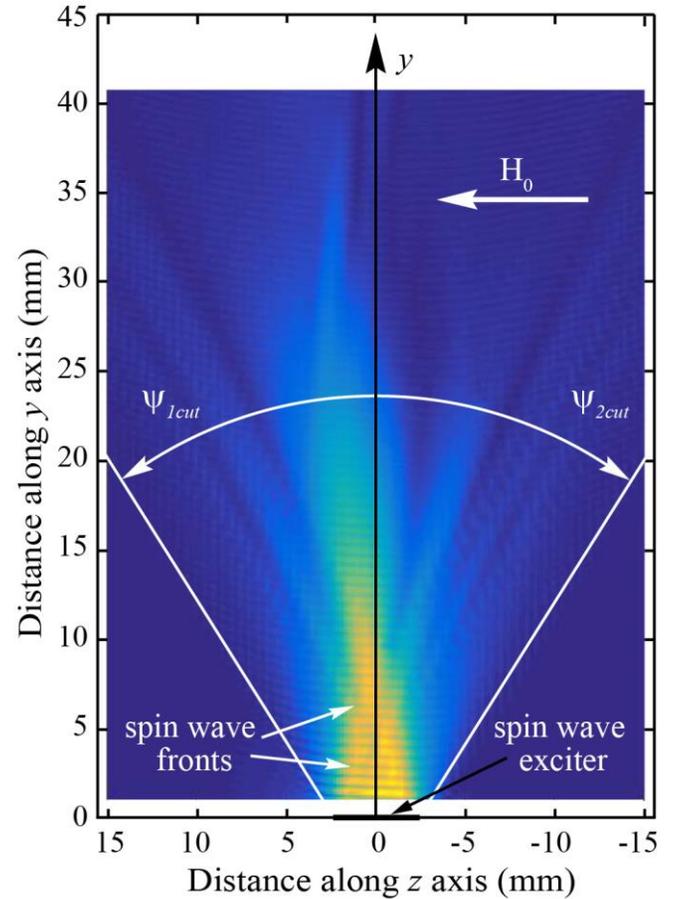

Fig. 5. Superposition of amplitude and phase distributions in the plane of ferrite film for the beam of spin wave with collinear vectors $\mathbf{k}$ and $\mathbf{V}$. Beam has the next parameters: $f_0 = 3242$ MHz, $k_0 = 110.8$ cm$^{-1}$, $\lambda_0 = 567$ μm, $\lambda_0/D = 0.113$, $\varphi_0 = 0°$, $\Delta\psi_{exp} = 12.9°$, $\sigma_{exp} = 2$. . Colour or greyscale change corresponds to the 3 dB change of spin wave amplitude relative to the maximal amplitude. Spin wave fronts with the same phase are shown also.

Comparison of parameters for spin wave beam in Fig. 4 and 5 shows that the beam in Fig. 4 almost does not change its width along the whole its trajectory ($\Delta\psi_{exp} = 0.4°$ and $\sigma_{exp} = 0.03$). Quite the contrary, spin wave beam in Fig. 5 is significantly expanded ($\Delta\psi_{exp} = 12.9^o$ and $\sigma_{exp} = 2$), despite the fact that for this beam the ratio $\lambda_0/D = 0.113$ is 2 times smaller than the ratio $\lambda_0/D = 0.222$ for the beam in Fig. 4. The experimental results are in good agreement with theoretical investigations, calculations and formulas in the work [1, 2].

In summary, we investigate experimentally the diffraction patterns of the surface spin wave excited by arbitrarily oriented linear transducer in tangentially magnetized ferrite film for the case where the transducer length $D$ is much larger than the wavelength $\lambda_0$. In the study there is used the scanning probe method, that give possibility to visualise the amplitude and phase distributions of the spin wave along the film surface. The angular beam width of spin waves is measured experimentally and is calculated theoretically by means of the general formula derived in [2] for the angular beam width in anisotropic media. It is proved experimentally that as a distinct from the beams in isotropic media the angular beam width $\Delta\psi$ of the surface spin wave is not *constant* value: it depends on transducer orientation $\varphi_0$ and can take values *greater* or *smaller* than the ratio $\lambda_0/D$ (where $\lambda_0$ is exited wavelength and $D$ is the exciter length). Moreover, it is found such experimental parameters and transducer orientation, at which angular beam width $\Delta\psi$ is about zero (it means physically, that the beam retains its absolute width during propagation). Thus it is shown that such phenomenon as "superdirectional propagation of the waves" exists in the nature. This phenomenon takes place when the wave vector orientation $\varphi_0$ (or the transducer orientation) corresponds to the inflexion point of isofrequency dependence for the excited spin wave. It is also evidently that well known Rayleigh criterion used in isotropic media can't be used to estimate the angular width of spin wave beams. An example of the surface spin wave diffraction gives us hope that the same diffractive phenomena can take place in the other anisotropic media and structures.

In addition it was found that linear transducer oriented parallel to external constant magnetic field **H₀** excites surface spin wave with wave vectors **k** and group velocity vector **V** which are only approximately directed along the perpendicular to the vector **H₀**, since there is a small phase difference between the initial and final points of linear transducer.

The experimental results are in good agreement with theoretical investigations, predictions, calculations and formulas on the basis of works [1, 2]. Moreover, on the basis of theory [1, 2] it was predicted another geometry in which superdirectional spin wave beam (with zero angular width) can be excited [4, 5].

The work is supported by Russian Foundation for Basic Research (project No. 17-07-00016).

* edwin@ms.ire.rssi.ru